\documentclass[12pt]{article}
\newsavebox{\ns}
\newsavebox{\dbrane}
\newsavebox{\dbshort}

\usepackage{epsfig}
\usepackage{amssymb}

\def\be{\begin{eqnarray}}
\def\ee{\end{eqnarray}}

\newcommand{\nn}{\nonumber}

\newcommand{\ft}[2]{{\textstyle\frac{#1}{#2}}}
\newcommand{\eqn}[1]{(\ref{#1})}

\def\Dslash{\,\,{\raise.15ex\hbox{/}\mkern-12mu D}}
\def\Dbarslash{\,\,{\raise.15ex\hbox{/}\mkern-12mu {\bar D}}}
\def\delslash{\,\,{\raise.15ex\hbox{/}\mkern-9mu \partial}}
\def\delbarslash{\,\,{\raise.15ex\hbox{/}\mkern-9mu {\bar\partial}}}
\def\pslash{\,\,{\raise.15ex\hbox{/}\mkern-9mu p}}
\def\calDslash{\,\,{\raise.15ex\hbox{/}\mkern-12mu {\cal D}}}

\begin{document}

\begin{titlepage}

\begin{center}
\today
\hfill hep-th/0202012\\
\hfill MIT-CTP-3238 \\

\vskip 1.5 cm
{\large \bf The Moduli Space of BPS Domain Walls}
\vskip 1 cm 
{David Tong}\\
\vskip 1 cm
{\sl Center for Theoretical Physics, 
Massachusetts Institute of Technology, \\ Cambridge, MA 02139, U.S.A.\\
{\tt dtong@mit.edu}}

\end{center}

\vskip 0.5 cm
\begin{abstract}
${\cal N}=2$ SQED with several flavors admits multiple, static 
BPS domain wall solutions. We determine the explicit 
two-kink metric and examine the dynamics of colliding domain 
walls. The multi-kink metric has a toric K\"ahler structure and 
we reduce the K\"ahler potential to quadrature. 
In the second part of this paper, we consider semi-local 
vortices on ${\bf R}\times {\bf S}^1$. We argue that, in the 
presence of a suitable Wilson line, the vortices 
separate into domain wall constituents. These play the role of 
fractional instantons in two-dimensional gauge theories and 
sigma-models.

\end{abstract}

\end{titlepage}

\pagestyle{plain}
\setcounter{page}{1}
\newcounter{bean}
\baselineskip16pt

\subsection*{Introduction}

The concept of the moduli space is one of the most important 
tools in the study of solitons. Originally introduced by Manton 
to describe the classical scattering of monopoles \cite{manton}, 
it is now appreciated that the topology and geometry of soliton 
moduli spaces also encodes many of the quantum properties of 
supersymmetric theories. Examples include the spectrum of 
solitonic bound states, and non-perturbative contributions to 
correlation functions. This has played a pivotal role in untangling 
the web of dualities in field and string theories. 

While the moduli spaces of instantons, monopoles and vortices 
have all been studied in detail, less attention has been paid 
to the moduli space of domain walls. Indeed, in most theories 
there is a force between widely separated domain walls 
\cite{nick,paul}\footnote{This remains true even when the domain 
walls are mutually BPS \cite{multikink}.}, and any attempt to 
describe the dynamics using a moduli space approximation 
requires the introduction of a potential \cite{man}. 
Nevertheless, there do exist 
models where the force between domain walls vanishes, resulting in 
a moduli space of static multi-kink solutions with arbitrary 
separation. These include a class of generalized Wess Zumino 
models \cite{shif,ilg}, ${\cal N}=1^\star$ theories 
\cite{boris}, and massive sigma models  
\cite{multikink}. 

In this paper we shall consider domain walls in ${\cal N}=2$ 
supersymmetric QED with a FI parameter and $N$ flavors of 
electrons\footnote{This 
is a slight generalization of the model considered in 
\cite{multikink}, reducing to it in the strong coupling 
limit \cite{witten}.}. If each 
electron has a different mass there are $N$ isolated vacua, 
implying the existence of BPS domain walls. As will be 
reviewed below, a generic domain wall decomposes into several 
``fundamental'' domain walls, each of which carries an independent 
position and phase collective coordinate. The moduli space of 
solitons is thus a toric K\"ahler manifold \cite{multikink}.

In fact, this theory has a much richer spectrum 
of solitons than one might guess through a naive homotopy 
group argument. As well as domain walls, there also exist 
superconducting BPS strings, which carry a global current 
\cite{ab}. Moreover, at least in the strong coupling limit, these 
strings can end on the domain wall where they are charged under a 
localized $U(1)$ gauge field  \cite{dbranewall}. (The gauge field 
is dual to the phase collective coordinate). 
In other words, this 
field theory provides a simple model of D-brane physics.

The paper is organized as follows. In the next section 
we introduce the abelian-Higgs model of interest. We review the 
first order domain wall equations, as well as the connection to the 
sigma-model kinks of \cite{multikink}. We then proceed to consider the 
metric on the moduli space of solutions and derive the K\"ahler 
potential in integral form (equation \eqn{kahler}). In the case of 
two kinks with identical masses, this integral can be performed (the 
resulting explicit metric is given in equation \eqn{2metric}) and 
some of the properties and implications of this metric 
are analyzed. In the second part 
of this paper, we change tack and discuss semi-local 
vortices on ${\bf R}\times {\bf S}^1$ with a Wilson line 
for the flavor symmetry group. Motivated by the 
analogy with instantons and monopoles \cite{leeyi}, we demonstrate that 
the vortices decompose into multiple kink solutions. This 
provides a mechanism for calculating fractional instanton 
effects in strongly coupled two-dimensional sigma models. 

A note on the quantum theory: in this paper we concentrate on 
domain walls in a $d=3+1$ dimensional abelian gauge theory. 
However, due to the existence of a Landau-pole, the theory is not 
well-defined at the quantum level. The same is true of the theory 
lifted to $d=4+1$ dimensions, which may be of interest in the 
context of brane-world scenarios\footnote{One cannot lift the 
theory with a mass gap to dimensions greater than $4+1$ while preserving 
supersymmetry.}. Since we restrict ourselves 
to classical aspects of domain walls, no harm is done. However, 
to find the same domain walls as quantum objects, we must 
consider the dimensional reduction of the theory to $d=2+1$ or 
$d=1+1$. Alternatively they appear as instantons in gauge quantum 
mechanics. In each of these cases, the metric on the moduli space 
remains the same.  

Finally, throughout the paper we stress the many similarities that 
exist between the domain walls and monopoles, as well as between 
semi-local vortices and Yang-Mills instantons. These similarities 
add to the growing evidence that there exists a quantitative 
correspondence relating these solitons \cite{dorey,ff}.

\subsection*{Gauge Theory Domain Walls}

Our starting point is $d=3+1$, ${\cal N}=2$ supersymmetric 
$U(1)$ gauge theory coupled to $N$ hypermultiplets. The bosonic 
part of the Lagrangian is given 
by,
\be
{\cal L}&=&\frac{1}{4e^2}F^2+\frac{1}{2e^2}|\partial\phi|^2 
+\sum_{i=1}^N\left(|{\cal D}q_i|^2+|{\cal D}\tilde{q}_i|^2 
\right) -\sum_{i=1}^N|\phi-m_i|^2(|q_i|^2+|\tilde{q}_i|^2) \nn\\ 
&& - \frac{e^2}{2}(\sum_{i=1}^N|q_i|^2-|\tilde{q}_i|^2
-\zeta)^2-\frac{e^2}{2}|\sum_{i=1}^N\tilde{q}_iq_i|^2
\label{lag}\ee
Each scalar field $q_i$ has charge $+1$ under the gauge group, 
while the $\tilde{q}_i$ have charge $-1$. Each pair is assigned  
a complex mass $m_i$ which may always be chosen to satisfy 
$\sum_im_i=0$. The complex scalar field $\phi$ lives in the 
vector multiplet and is neutral under the gauge group. 
Finally, we require that $\zeta$, the real FI parameter appearing 
in the D-term, is non-zero. This ensures the theory lies 
in its Higgs phase. Without loss of generality\footnote{A 
possible complex FI parameter which would 
appear in the F-term in \eqn{lag} has been set to zero using 
the $SU(2)_R$ R-symmetry of the action.} we set $\zeta>0$. 

For vanishing masses the theory enjoys a $SU(N)$ flavor symmetry 
and a moduli space of vacua given by $T^\star{\bf CP}^{N-1}$.  
In this paper we will be interested in the case of non-zero,  
distinct masses: $m_i\neq m_j$ for $i\neq j$. This breaks the  
flavor symmetry to the maximal torus $U(1)^{N-1}$ and lifts 
all but $N$ isolated vacua, lying on the zero section of 
$T^\star {\bf CP}^{N-1}$,
\begin{center}
{ Vacuum $i$:}\quad $\phi=m_i\ \ ,\ \ |q_j|^2=\zeta\delta_{ij}\ \ ,
\ \ |\tilde{q}_j|^2=0$
\end{center}
For generic masses $m_i$, there exist BPS domain wall solutions 
interpolating between any given pair of vacua. However, in 
order to find a moduli space of domain walls we need to 
restrict the mass parameters to be real\footnote{This is 
entirely analogous to the situation with monopoles in higher rank gauge 
groups, in which a moduli 
space only exists if the vacuum expectation value is real \cite{tim}.}
: ${\rm Im}\,(m_i)=0$. This immediately leads to the important 
corollary that there is a natural ordering to the vacua. We 
choose the ordering $m_{i+1}<m_i$ for all $i$. 

Since certain fields will not appear in the domain wall solutions 
discussed below, we set them to zero at this stage:
\be
{\rm Im}\,(\phi)=\tilde{q}_i=F=0
\label{set}\ee
Their sole role was to complete the supersymmetry multiplets, and to 
cancel a potential gauge anomaly. (In fact the field strength will be 
resurrected below when we come to discuss dynamics). 
In particular, from now 
on the field $\phi$ will always be assumed to be purely real. 
We choose the domain wall to lie in the $(x^2-x^3)$ plane, 
so that the only non-zero space-time field variations are 
in the $x\equiv x^1$ direction. We write $\partial\equiv\partial_1$. 
The BPS equations, first derived in \cite{kinky}, 
can be determined by simply completing the square in the 
Hamiltonian,
\be
{\cal H}&=&\frac{1}{2e^2}(\partial\phi)^2+\sum_{i=1}^N
|{\cal D}q_i|^2+\sum_{i=1}^N(\phi-m_i)^2|q_i|^2 +
\frac{e^2}{2}(\sum_{i=1}^N|q_i|^2-\zeta)^2 \nn\\
&=& \frac{1}{2e^2}(\partial\phi\mp e^2(\sum_{i=1}^N |q_i|^2 
-\zeta))^2 + \sum_{i=1}^N|{\cal D}q_i\mp(\phi-m_i)q_i|^2 \pm T
\nn\ee
For the kink interpolating between the $i^{\rm th}$ vacuum 
at $x\rightarrow -\infty$ and the $j^{\rm th}$ vacuum 
at $x\rightarrow +\infty$, the topological charge $T$ 
is given by,
\be
T=\left[\sum_{i=1}^N(\phi-m_i)|q_i|^2-\phi\zeta
\right]^{+\infty}_{-\infty}=\zeta(m_i-m_j)
\label{t}\ee
where we have chosen $j>i$ which requires use of the upper signs 
in the Hamiltonian.  The Bogomoln'yi equations are 
therefore given by,
\be
\partial\phi&=&e^2(\sum_{i=1}^N|q_i|^2-\zeta) \label{bog1} \\ 
{\cal D}q_i&=&(\phi-m_i)q_i
\label{bog2}\ee
It is simple to show that for $j<k<i$, these Bogomoln'yi 
equations require $q_k\equiv 0$. For this reason, we now restrict 
attention to the maximal domain wall interpolating between 
the $1^{\rm st}$ and $N^{\rm th}$ vacua, which has tension  
$T=\zeta(m_1-m_N)$. Any other domain wall may be embedded maximally 
in a theory with fewer flavors. 

The second Bogomoln'yi equation \eqn{bog2} is easily  
solved,
\be
q_i=\sqrt{\zeta}\exp\left(\psi-m_i(x-x_0)-
\sum_{a=1}^{N-2}\alpha_i^ar_a\right)
\label{sol}\ee
where $\alpha$ is a fixed, rank $(N-2)$ real matrix satisfying 
$\sum_i\alpha_i^a=\sum_im_i\alpha_i^a =0$, and the complex 
function $\psi$ is determined by,
\be
\partial\psi=\phi+iA
\nn\ee
with $A\equiv A_1$ the gauge potential. By \eqn{set}, the imaginary 
part of $\psi$ is pure gauge. It may be set to zero when considering 
static solutions, but will play an important role when we turn to the 
dynamics of domain walls. Most important in the solution \eqn{sol} 
are the putative collective coordinates. These are the center of 
mass $x_0$ and the parameters $r_a$, $a=1,\ldots,N-2$ which are related 
to the separation of neighbouring domain walls. Each is complex, 
with real and imaginary parts,
\be
x_0=X_0+i\theta_0\quad,\quad r_a=R_a+i\theta_a
\label{complex}\ee
and will provide $N-1$ complex coordinates on the domain wall 
moduli space. When $m_i$ and $\alpha^a_i$ are rational, the 
corresponding $\theta$ is periodic. In contrast, when 
$m_i$ and $\alpha^a_i$ are irrational, $\theta\in {\bf R}$. 
Note that there exists some ambiguity in fixing the 
matrix $\alpha^i_a$ which is related to the possiblity of 
performing coordinate redefinitions on the moduli space. This 
ambiguity may be naturally removed by insisting that, asymptotically, 
the parameters $R$ coincide with the relative separations of 
far-separated domain walls. We shall do this explicitly for the two-kink 
metric but in general it remains an open problem.
 
However, we must not be too hasty in concluding 
that multi-domain wall solutions exist, since we have still 
to satisfy the first Bogomoln'yi equation \eqn{bog1} which 
now reads,
\be
\frac{1}{\zeta e^2}\,\partial^2{\rm Re}\,(\psi)=\sum_{i=1}^N
\exp\left(2{\rm Re}\,(\psi)-2m_i(x-X)-2\alpha_i^aR_a\right)-1
\label{biggy}\ee
Note that we have left the sum over $a=1,\ldots,N-2$ implicit 
in this equation, and shall continue to do so for the remainder 
of the paper. This non-linear, somewhat unpleasant, differential equation, 
which defines ${\rm Re}\,(\psi)$ as a function of the real 
variables $(x-X)$ and $R_a$, is further complicated by 
the boundary conditions,
\be
{\rm Re}\,(\psi)\rightarrow\left\{\begin{array}{ccc} 
m_1(x-X)+\alpha_1^aR_a & \quad & x\rightarrow -\infty \\ 
m_N(x-X) +\alpha_N^aR_a & \quad & 
x\rightarrow +\infty\end{array}\right.
\nn\ee
I do not know if solutions exist for all values of the dimensionful  
parameter $\zeta e^2$. However, it is possible to write down 
a formal solution as a perturbative series in the dimensionless 
parameter $e^{-2}$. The strong coupling expansion takes the form, 
\be
{\rm Re}\,(\psi)=\sum_{p=0}^\infty \frac{1}{e^{2p}}\psi_p\ .
\label{series}\ee
Then, in the strict strong coupling limit $e^2\rightarrow \infty$, 
the solution is
\be
\exp(2\psi_0)=\left(\sum_{i=1}^N\exp(-2m_i(x-X)-2\alpha_i^aR_a
)\right)^{-1}
\label{psi0}\ee
which indeed has the correct boundary conditions. This may be 
understood as the long-wavelength approximation to the true 
solution to \eqn{biggy}. The remaining 
$\psi_p$ for $p>1$ are determined in an iterative fashion 
by the equation,
\be
\partial^2\sum_{p=0}^\infty\frac{1}{e^{2{(p+1)}}}\psi_p 
=\zeta\sum_{n=1}^\infty\frac{1}{n!}\left(\sum_{p=1}^\infty 
\frac{2}{e^{2p}}\psi_p\right)^n
\nn\ee
which ensures that $\psi_p\rightarrow 0$ as $x\rightarrow \pm\infty$, 
so the boundary conditions are preserved. 
Thus, there exist solutions to \eqn{biggy} enjoying the full 
compliment of $N-1$ complex collective coordinates, 
at least in a neighborhood of $e^{-2}=0$. The size of this 
neighborhood 
is determined by the radius of convergence of the sum \eqn{series}, 
given by the limit $|\psi_p/\psi_{p+1}|$ as $p\rightarrow \infty$. 
It would be interesting to determine whether the solutions exhibit 
a phase transition as the coupling constant is varied, 
or whether the radius of convergence is infinity or (more  
disappointing) zero.

The strong coupling limit $e^2\rightarrow\infty$, which played 
an important role in determining the existence of the 
solution, is familiar from linear sigma-models \cite{witten}. 
From the expression for the scalar potential 
\eqn{lag}, it is clear that this limit restricts us 
to the Higgs branch ${\cal V}=T^\star{\bf CP}^{N-1}$. 
The presence of mass terms for the hypermultiplets 
induces a potential on the Higgs branch which, by supersymmetry 
requirements, is proportional to the length of a tri-holomorphic 
Killing vector on ${\cal V}$ \cite{agf}. This Killing vector 
is determined by the global flavor symmetry preserved by the 
masses $m_i$. Domain walls in such massive sigma-models 
have been extensively studied in the literature and, 
in particular, the existence of multi-kink solutions  
was demonstrated in \cite{multikink} using both Morse theory as 
well as more direct techniques. It was further shown in 
\cite{multikink} that the coordinates $R_a$ do indeed 
parameterize the separations between ``fundamental kinks'', 
each of which interpolates from the $i^{\rm th}$ vacuum  
to the $(i+1)^{\rm th}$ vacuum. The maximal kink considered 
here is comprised of $(N-1)$ fundamental kinks. A similar 
decomposition of kinks has also been found in $SU(N)\times {\bf Z}_2$ 
gauge models \cite{pogosian}.

Moreover, the massive sigma-models also admit a wide range of 
BPS solutions including 
intersecting domain walls \cite{intdom}, string-lumps \cite{qlump}, 
strings ending on domain walls \cite{dbranewall} and 
dyonic domain walls \cite{qkink}. It seems likely that each 
of these has a generalization to the finite coupling gauge 
theory considered here. This is certainly true of strings, 
as shown in \cite{ab}, and it a simple exercise to generalize 
the BPS equations of all these solitons to the gauge theory 
context.

\subsection*{The Moduli Space of Domain Walls}

We turn now to the dynamics of domain walls. We work in 
the moduli space approximation; in other words, we consider 
solutions to the static BPS equations. The 
complex collective coordinates are then promoted to fields 
on on the domain wall world-volume,  
$x_0(\xi)$ and $r_a(\xi)$. For small fluctuations, the low-energy 
dynamics of the 
soliton is described by a $d=2+1$ dimensional sigma model 
with target space toplogically of the form \cite{multikink}
\be
{\cal M}_{N-1}={\bf R}\times\frac{{\bf R}\times \tilde{\cal M}_{N-1}}{{\cal G}}
\nn\ee
where the first two ${\bf R}$ factors parameterize the center-of-mass 
and overall phase of the soliton respectively. Newton's third law 
ensures each is endowed with a flat metric. 
As with monopoles, motion along the second ${\bf R}$ 
factor recovers the dyonic domain walls discussed in \cite{qkink}. 
All interesting dynamics is 
encoded in the metric on $\tilde{\cal M}_{N-1}$, the relative kink moduli 
space, which has complex dimension $(N-2)$. This inherits both a complex 
structure from \eqn{complex} as well as $(N-2)$ holomorphic $U(1)$ isometries 
from the abelian flavor symmetries, and is thus a toric K\"ahler manifold. 
The quotient by the discrete group ${\cal G}$ acts only on the 
toric fibers. For generic domain wall masses, ${\cal G}={\bf Z}$. 
For certain, rational, choices of masses the second 
${\bf R}$ factor collapses to ${\bf S}^1$, and ${\cal G}$ is 
a finite group. This is identical to monopoles in higher rank 
gauge groups. Unlike monoples however, the symmetries of the 
problem allow the metric to contain constant cross terms between 
the center-of-mass and relative motion factors --- more on these 
later. 

To find the metric, we firstly study fluctuations around the 
domain wall background. We concentrate on variations with 
respect to time, but the final answer may be trivially extended to 
have $d=2+1$ Lorentz symmetry on the domain wall world volume. 
The linearized Bogomoln'yi equations are,
\be
\partial\dot{\phi}&=&e^2\sum_{i=1}^N(\dot{q}_i{q}^\dagger_i+q_i
\dot{q}^\dagger_i) \label{lbog1}\\
{\cal D}\dot{q}_i-i\dot{A}q_i&=&(\phi-m_i)\dot{q}_i+\dot{\phi}q_i
\label{lbog2}\ee
which are augmented by Gauss' law, determining the electric 
field $E=F_{01}$,
\be
\partial E=ie^2\sum_{i=1}^N(q_i{\cal D}_0{q}^\dagger_i
-{q}^\dagger_i{\cal D}_0q_i)
\label{gauss}\ee
This equation requires $E\neq 0$. However, it may be shown 
that neither ${\rm Im}(\phi)$ nor $\tilde{q}_i$ have zero modes, 
and so we continue to ignore them as per ansatz 
\eqn{set}\footnote{This is not true of the corresponding 
fermions and the superpartners of $\tilde{q}_i$ do yield fermionic 
zero modes.}. We choose to work in $A_0=0$ gauge, in which case the three 
equations above combine to give,
\be
\partial^2\left(\frac{\dot{q}_i}{q_i}\right)=
\partial(\dot{\phi}+i\dot{A})=2e^2\sum_{j=1}^N\dot{q}_jq_j^\dagger
\nn\ee
which is valid for all $i$ such that $q_i\neq 0$. 
The metric 
on the moduli space is, as usual, inherited from the 
kinetic terms in the action. After employing the 
above time-evolution equations, this can be brought into the 
form,
\be
{\cal L}&=&\int\ dx\ \frac{1}{2e^2}|\dot{\phi}+i\dot{A}|^2+\sum_{i=1}^N
\dot{q}_i\dot{q}_i^\dagger \nn\\
&=& \int\ dx\ \sum_{j=1}^N\dot{q}_j\dot{q}_j^\dagger-\frac{\dot{q}_i}{q_i}
\dot{q}_j^\dagger q_j\ \ \ \ \ \ \ \forall\ i \ (q_i\neq 0) \nn\\
&=& \int\ dx\ \sum_j|q_j|^2(\dot{\psi}+m_j\dot{x}_o-\alpha_j^a\dot{r}^a)
(m_j\dot{x}_0^\dagger-\alpha_i^a\dot{r}_a^\dagger)
\nn\ee
where, in the last equality, we have used the explicit expression 
for $q_j$ given in equation \eqn{sol}. To make further progress, 
we must find an expression for $\dot{\psi}$ in terms of the collective 
coordinates $x_0$ and $r_a$. This involves solving equation 
\eqn{biggy} which is currently beyond our reach apart from in 
the  strong coupling $e^2\rightarrow \infty$ limit. From now 
on, we therefore restrict ourselves to this regime of parameter 
space so that, using \eqn{psi0}, Gauss' law becomes
\be
\dot{\psi}_0=\frac{\sum_{i=1}^N(m_i\dot{x}_0+\alpha_i^a\dot{r}_a)
\exp(-2m_i(x-X)-2\alpha_i^aR_a)}{\sum_{j=1}^N\exp
(-2m_j(x-X)-2\alpha_j^aR_a)}
\nn\ee
Inserting this into the Lagrangian, and performing some of the 
more simple integrals, we find the metric on the ${\cal G}$-fold 
cover of ${\cal M}_{N-1}$ is given by,
\be
{\cal L}= \ft12 T|\dot{x}_0|^2 + 
\zeta(\alpha_N^a-\alpha_1^a)\dot{x}_o\dot{r}_a^\dagger 
+ {\rm h.c.} + \frac{\partial^2{\cal K}}{\partial R_a\partial R_b}
\dot{r}_a\dot{r}_b^\dagger
\label{metric}\ee
The cross-term between the center-of-mass $x_0$ and the relative 
separations $r_a$ reflects the fact that the generic domain wall 
solution breaks parity, so that the metric is not invariant under 
$x_0\rightarrow - x_0$.\footnote{Such terms are 
absent for higher co-dimension solitons (vortices, monopoles, 
instantons) by rotational symmetry. I thank Adam 
Ritz for a discussion on this issue.}. In relativistic 
terminology, the metric is ``stationary'', as opposed to 
``static'', in the $x_0$ coordinate. The interesting 
dynamics is contained within the K\"ahler potential 
${\cal K}(R_a)$ which is given by the integral,
\be
{\cal K}=\frac{\zeta}{4}\int dx\ \log\left(\sum_{i=1}^N\exp(-2m_ix-2
\alpha_i^aR_a)\right)
\label{kahler}\ee
Note that although this integral is divergent, all such terms 
are at most linear in $R_a$, and so do not contribute 
to the metric. The toric K\"ahler structure of the metric, which 
is required by the global and super- symmetries of the theory, 
is manifest in these coordinates. In particular, the K\"ahler 
structure ensures that the bosonic Lagrangian \eqn{metric} 
enjoys a supersymmetric extension. Since the domain walls 
are half BPS, their three-dimensional worldvolume dynamics 
preserves ${\cal N}=2$ supersymmetry (or four-supercharges). The 
relevant sigma-model 
was found long ago \cite{dan} and includes a four-fermi 
term coupling Grassmannian zero modes to the Riemann tensor 
associated with the metric 
\eqn{kahler}.

As mentioned in the introduction, the function ${\cal K}$ contains 
information 
about the classical scattering of domain walls, as well as 
quantum effects in lower dimensional theories. For 
example, we may dimensionally reduce the theory \eqn{lag} 
to $d=1+1$, with ${\cal N}=(4,4)$ 
supersymmetry. The spectrum of solitons is then determined by 
normalisable, harmonic forms on ${\cal M}_k$. The holomorphic 
subset of these forms survive as states in the ${\cal N}=(2,2)$ 
theory (in which the $\tilde{q}_i$ and their superpartners are 
removed). Alternatively, 
if we reduce further to $d=0+1$ quantum mechanics, the integral 
of the Euler form over ${\cal M}_k$ yields the $k$-instanton contribution 
to the four-fermi correlation function. 

\subsubsection*{\rm{\em The Two-Kink Metric}}

Let us now restrict attention to the simplest case: 
two kinks with equal tension $T=\ft12 M\zeta$. This occurs for the 
$N=3$ model with the parameters, 
\be
m_i=(\ft12 M,0,-\ft12 M)\quad,\quad \alpha_i=\ft16(\ft12 M,-M,\ft12M)
\label{machoice}\ee
With this choice, the system is symmetric under 
$x_0\rightarrow - x_0$, ensuring that the $d{x}_0 d{r}^\dagger$ 
cross-terms in the metric vanish. The moduli 
space takes the form,
\be
{\cal M}_2={\bf R}\times \frac{{\bf S}^1\times \tilde{\cal M}_2}{{\bf Z}_2}
\nn\ee
The periodicities of the two angular variables can be found by careful 
examination of the solutions \eqn{sol}. The ${\bf S}^1$ factor is 
parameterized by 
$\theta_0\in [0,4\pi/M)$. The relative moduli space is parameterized by the 
collective coordinates $R\in{\bf R}$ and $\theta\in[0,8\pi/M)$. 
The ${\bf Z}_2$ symmetry acts only on these periodic variables, 
\be
{\bf Z}_2: \theta_0\rightarrow \theta_0+2\pi/M ,\quad\theta\rightarrow 
\theta+4\pi/M
\nn\ee
Most importantly, with the choice of parameters \eqn{machoice} the 
K\"ahler potential 
given in equation \eqn{kahler} simplifies tremendously; in fact, to 
the point where Mathematica is happy to perform the integral with 
minimal complaint. We find the relative moduli space metric to be, 
\be
ds^2=\ft1{16} M\zeta\, F(R)\,(dR^2+d\theta^2)
\label{2metric}\ee
where all interactions are encoded in the function $F$;
\be
F(R)&=&e^{MR/2}\int dy\ \frac{e^{-y}+e^y}{(e^{-y}+e^y+e^{MR/2})^2} \nn\\ 
&=&\frac{2e^{MR/2}}{(e^{MR}-4)^{3/2}}\left[e^{MR/2}
(e^{MR}-4)^{1/2}+4\log\left(\frac{2+e^{MR/2}-(e^{MR}-4)^{1/2}}
{2+e^{MR/2}+(e^{MR}-4)^{1/2}}\right)\right]
\nn\ee
Note that, despite appearances to the contrary, the function $F$ 
is real and positive definite. The apparent singularity at 
$e^{MR}=4$ is quite illusory: $F$ is smooth at that point with 
value $4/3$. As mentioned in the introduction, this 
function contains information about the spectrum of 
domain walls in $d=1+1$ dimensional gauge theories, as 
well as instanton effects in gauge quantum mechanics. Here 
we merely extract some simple physics concerning the classical 
dynamics. Firstly, let us consider the 
limit of far-separated domain walls. As 
$R\rightarrow\infty$, the metric becomes
\be
ds^2 \rightarrow \ft18 M\zeta \left(1 - (2MR-4)e^{-MR} 
+{\cal O}(e^{-2MR})\right)\,(dR^2+d\theta^2)
\nn\ee
The constant term, with the factor of $M/8$ shows that we have 
correctly identified $R$ as the large-distance separation between 
the kinks (this can be traced back to the choice of normalization of 
$\alpha$ in \eqn{machoice}). The leading order correction to free 
motion is exponentially suppressed as  expected for any soliton  
in a theory with a mass gap. The long-range velocity dependent 
force is,
\be
\ddot{R}=-e^{-MR}(M^2R-3M)(\dot{R}^2-\dot{\theta}^2)
\nn\ee
For suitably large $\dot{\theta}$, the two kinks repel. 
If, instead, the kinks have constant relative phase, 
the overall minus sign implies that the velocity dependent force 
is attractive at large separation. In fact, numerical studies show 
that this attractive force persists for all values of $R$. Thus 
kinks moving initially apart will continue on such a trajectory, 
slowing but not halting. In contrast, kinks moving towards 
each other will increase their speed. Assuming the velocities 
remain small so that the moduli space approximation is 
valid, we can determine their fate by examining the metric 
as they approach. In the limit $R\rightarrow -\infty$, the 
function $F\rightarrow \ft12\pi\exp(MR/2)$. After changing to 
coordinates
\be
L=\left(\frac{\pi\zeta}{2M}\right)^{1/2}\exp(MR/4)\quad\in{\bf R}^+
\nn\ee
then, as $L\rightarrow 0$, the metric becomes
\be
ds^2 \rightarrow dL^2 + \ft1{16} M^2 L^2 d\theta^2
\nn\ee
which, for the specific range $\theta\in [0,8\pi/M)$, is 
non-singular. Thus the two-kink moduli space is smooth, 
the collision is elastic and the domain walls rebound with 
their phases exchanged, $\theta\rightarrow \theta+4\pi/M$.

\subsection*{Fractional Vortices}

In this section, we discuss the relationship between the 
gauge theory domain walls considered above, and BPS 
semi-local vortices \cite{semi,ruiz} (for a review, 
see \cite{av}). The latter solitons are vortices in an 
abelian gauge theory with a multi-component Higgs field. 
This is precisely the model of equation \eqn{lag}  if 
we set the mass terms $m_i$ to zero. The vortices 
have $\tilde{q}_i=\phi=0$, but a non-zero 
magnetic field, say $B=F_{13}$ for a vortex string 
extended in the $x_2$ direction. The BPS equations are,
\be
B&=&e^2(\sum_{i=1}^N|q_i|^2-\zeta) \nn\\ 
{\cal D}_1q_i&=&{\cal D}_3q_i 
\nn\ee
Suppose we dimensionally reduce on the $x_3$ direction, 
so that $\partial_3\equiv 0$ on all dynamical fields, and 
we further rename $A_3\equiv {\rm Re}(\phi)$. Then the vortex 
equations are precisely the domain wall equations \eqn{bog1} and 
\eqn{bog2} if, upon dimension reduction, we impose a Wilson line 
for the $SU(N)$ flavor symmetry, introducing the masses $m_i$.

This situation is reminiscent of the relationship between 
monopoles and instantons. In this case, dimensional 
reduction of the self-dual Yang-Mills equations, with a Wilson 
line for the $SU(N)$ gauge symmetry, results in the BPS 
monopole equation. In fact, in the monopole case, there is 
a deeper relationship between instantons and monopoles, 
discovered by Lee and Yi \cite{leeyi}. These authors consider 
$SU(N)$ instantons compactified on 
${\bf R}^3\times {\bf S}^1$  with a Wilson line around the 
${\bf S}^1$, breaking the gauge group to the maximal torus. 
Such configurations are known as calorons. 
The dimension of the moduli space of calorons is the same 
as that of instantons in flat space; for Pontyagrin number 
$k$, there are $4kN$ moduli. 
Lee and Yi show that these collective coordinates may be 
understood as the position and internal phase of $kN$ 
magnetic monopoles of $N$ different types. Recall that on 
flat ${\bf R}^3$, an $SU(N)$ gauge theory plays host to only 
$(N-1)$ different types of monopoles, one for each simple 
root of the Lie algebra. For the caloron, the extra monopole 
is associated to the affine root of the Lie algebra, and 
arises because of the periodic nature of the Higgs field. The 
simplest way to see this result is using the string set-up of 
D0-D4-branes compactified on a circle. 

In the remainder of this paper, we would like to show that 
a similar phenomenon happens for semi-local vortices. Related 
observations were also made in \cite{qkink}. As in 
the case of instantons and monopoles, the simplest way to 
see this result is through a brane construction. In fact, 
we choose to model the $d=2+1$ theory which is simply the 
dimensional reduction of the model considered up to 
now\footnote{The 
$d=3+1$ dimensional theory has a Landau pole. This doesn't 
affect the classical discussion of this paper but, due to 
the omniscience of string theory, makes a brane discussion 
more subtle.}, which we subsequently compactify on 
${\bf R}^{1,1}\times {\bf S}^1$. 
There are (at least) two ways to construct 
semi-local 
vortices in string theory, and each has its advantages. 
The first method uses IIA string theory with a background of 
D2- and D6-branes, together with a NS-NS $B_2$-field,
\be
D2:&& 012 \nn\\
N\times D6:&& 0123456 
\nn\ee
with anti-self dual $B_{\mu\nu}$ for $\mu,\nu=3,4,5,6$. 
The theory on the D2-brane is the $U(1)$ gauge theory 
of interest, with $\zeta\sim |B|$. In this set-up, the 
Chern-Simons terms on the D2-brane world-volume imply 
that the semi-local vortex may be thought of as a D0-brane 
absorbed within the D2-brane. Unfortunately, moduli counting 
is difficult to see from this perspective, as the D0-brane 
does not preserve supersymmetry when it is removed from the 
D2-D6-bound state. Nevertheless, we shall have use for this 
construction later.

\begin{figure}
\begin{center}
\epsfxsize=6in\leavevmode\epsfbox{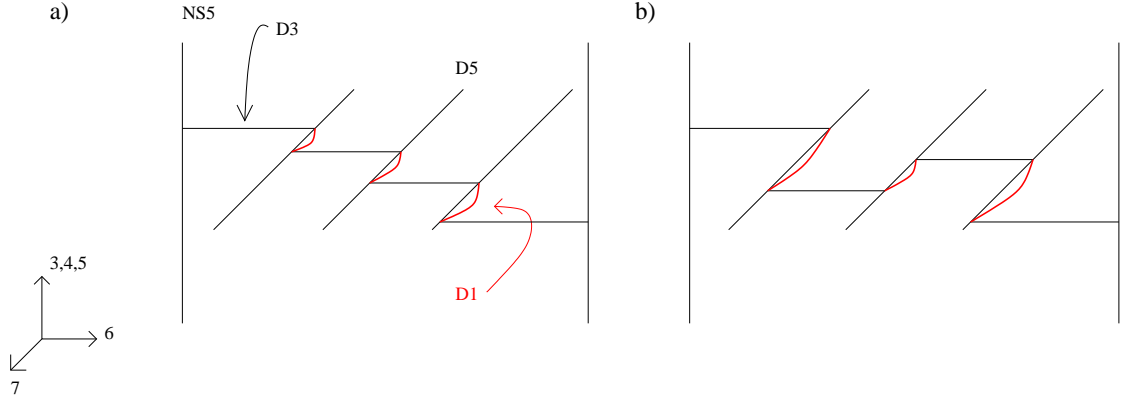}
\end{center}
\caption{Semi-local vortices from IIB branes. Figure 1a) has 
$\tilde{q}_i=0$, and finite mass BPS states exist. In Figure 1b, 
$\tilde{q}_i\neq 0$ and BPS states do not exist. The middle D-string 
has opposite orientation to the others, and breaks supersymmetry.}
\label{wall1}
\end{figure}

The second construction of semi-local vortices uses the 
IIB Hanany-Witten set-up,
\be
2\times NS5:&& 012345 \nn\\
N\times D5: && 012789 \nn\\
D3: && 0126 \nn\\
k\times D1: && 07
\nn\ee
The full configuration is drawn in Figure 1. The low-energy 
theory on the D3-brane is the $d=2+1$, ${\cal N}=4$ $U(1)$ 
gauge theory with $N$ flavors described in \eqn{lag}
Although this set-up employs more branes, it has the advantage 
that all parameters have geometric origins. For example, the 
real FI parameter is given by the relative positions 
of the NS5-branes,
\be
\left.x^8\right|_{NS5}=\left.x^9\right|_{NS5}=0\quad ,
\quad \left.\Delta x^7\right|_{NS5}=\zeta
\nn\ee
while the complex masses are related to the relative 
positions of the D5-branes in the $x^4$ and $x^5$ directions. 
For the purposes of discussing vortices, we set the masses to zero. 
The vevs of $q_i$ and $\tilde{q}_i$ are determined by the positions 
of the D3-brane segments suspended between the D5-branes. We denote 
the $(N+1)$ segments as $D3_\alpha$. For BPS 
vortices to exist, we require $\tilde{q}_i=0$. In the brane 
picture, this translates to 
\be
\left. x^8\right|_{D3_\alpha}=\left. x^9\right|_{D3_\alpha} = 0\quad,
\quad \left. x^7\right|_{D3_{\alpha+1}} > \left. x^7\right|_{D3_\alpha}
\nn\ee
In this case, $k$ parallel D-strings, lying in the $x^7$ 
direction, may connect the first D3-brane segment with the last, 
while preserving supersymmetry ---  see Figure 1a. This is the 
semi-local vortex with magnetic charge $k$. Each of 
these D-strings splits into $N$ separate segments as illustrated 
in the picture. Since each of these segments is free to roam 
the $(x^1-x^2)$ plane, the moduli space of semi-local vortices 
has real dimension $2kN$. This is counting is indeed correct, as 
shown in \cite{ruiz}. Moreover, this provides the first piece of 
numerological evidence that, as for instantons and monopoles, 
the charge $k$ semi-local vortex may decompose into $kN$ parallel 
domain walls when placed on a circle.

\begin{figure}
\begin{center}
\epsfxsize=5in\leavevmode\epsfbox{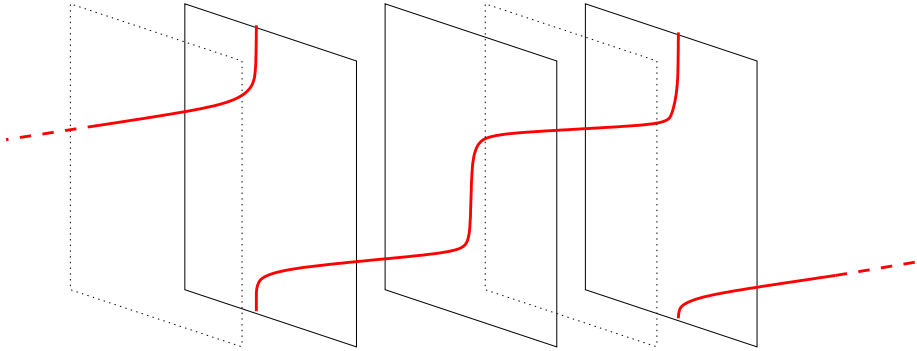}
\end{center}
\caption{Kinks as fractional vortices: the $N=2$, $k=1$ model. 
The infinite array of branes is periodic mod 2. The 4 
collective coordinates of the vortex are seen as 
the positions and phases of two kinks.}
\label{wall2} 
\end{figure}

One may wonder why a single segment of D-string does not qualify 
as a BPS state of the theory. To see this, note that the flux 
from such a D-string is deposited on the D3-brane segment, where 
it spreads out. Since the D3-brane is non-compact only in 
two spatial dimensions, this state has a logarithmically 
divergent mass. Only when the D-strings form a closed, oriented 
path from the first D3-brane to the last can this flux escape onto 
the NS5-branes, leading to a state of finite 
energy. Note also that the brane picture suggests that 
finite mass states also exist when $\tilde{q}_i\neq 0$ --- 
see Figure 1b. However, since the D-strings are not 
parallel in this case, the state breaks supersymmetry.

We turn now to the question of vortices on 
${\bf R}\times {\bf S}^1$, where the circle is taken to 
have radius $R$. To realize this in string theory, we may work 
with either of the brane constructions above, and make $x^2$ 
periodic. Since the IIA set-up is the less complicated (and 
easier to draw) we choose to work in the D2-D6-system. If a 
Wilson line 
for the $SU(N)$ flavor symmetry is introduced, upon 
T-duality the D6-branes become D5-branes separated in the 
$x^2$ direction. The flat D2-brane becomes a D-string oriented 
in the $x^1$ direction. The result is the familiar D1-D5 system. 
The presence of the $B_2$-field, which is not affected by 
T-duality, induces a force between the D-string and the 
D5-branes. This reproduces the $N$ vacuum states, in which 
the D-string lies within a given D5-brane. 
What becomes of the vortex? At the position of the 
vortex, the D2-brane is wrapping an internal cycle and 
thus, locally, is not extended in the $x^2$ direction. Upon 
T-duality, we therefore expect a D-string wrapping the dual 
$S^1$. The final configuration for $N=2$ is shown in Figure 2. 
The D-string interpolates to one of the vacuum states at 
$x_1\rightarrow \pm \infty$, lying within one of the D5-branes. 
However, at the location of the vortex, it leaves its asymptotic 
location to wind $k$ times around the circle ($k=1$ is drawn in the 
picture). On the way around, it may make a temporary home in any one 
of the other $(N-1)$ D5-branes it encounters. The 
existence of such BPS kinky D-string configurations was predicted in 
\cite{bt}, and further examined in \cite{kinky} where it was shown that 
they indeed correspond to the gauge theory domain walls discussed 
in this paper. 

Let us discuss the $N=2$ field theory in more detail. This theory 
only two vacua, located at $\phi=\pm m$. 
(Since the mass $m$ arose from a flavor Wilson line, it is periodic 
so we must have $m<\pi/R$). In 
flat space, this would imply the existence of a single 
kink, in which $\phi$ monotonically increases from $\phi=-m$ to 
$\phi = +m$. There is also a corresponding anti-kink 
in which $\phi$ decreases, in the other direction. However, 
as the brane picture clearly demonstrates, when placed 
on ${\bf R}\times {\bf S}^1$, there exists a further 
{\em kink} (as opposed to anti-kink) solution which interpolates 
from $\phi=+m$ to $\phi=-m$. This preserves the same supersymmetry 
as the original kink and is possible because $\phi$ arises 
from a Wilson line, $\phi=\int_{{\bf S}^1} A$. Invariance under 
large gauge transformations means that $\phi$ has period 
$2\pi/R$. This allows $\phi$ to interpolate from $+m>0$ 
to $-m<0$ with $\phi^\prime>0$ at all times. 
For the theory on the circle, the mass of the 
original kink is $2\tilde{\zeta}m$, where 
$\tilde\zeta=2\pi R\zeta$. In the limit $R\rightarrow 0$, we 
keep $\tilde{\zeta}$ fixed, to ensure that this kink 
remains in the spectrum. In contrast, the new kink which interpolates from 
$+m$ to $-m$, has mass $\tilde{\zeta}(2\pi/R-2m)$. As 
$R\rightarrow 0$, with $\tilde{\zeta}$ fixed this 
kink decouples, and we recover the situation described at 
the beginning of this paper. 

A periodic $\phi$ allows for the 
possibility of multiple domain wall solutions for theories with 
only two, or indeed one, vacuum states.  It would be interesting to 
examine the metric on these moduli spaces in more detail, 
especially in light of the fact that they give a K\"ahler 
deformation of the (semi-local) vortex moduli space. 

As stressed above, the preceding discussion is entirely analogous 
to that of instantons and monopoles given in \cite{leeyi}. In that 
case, the calorons have found an interesting application 
in supersymmetric quantum field theories. It has long been known 
that certain strongly coupled four-dimensional 
field theories receive non-perturbative contributions that 
have the characteristics of fractional instantons. The 
decomposition of instantons into 
monopoles when compactified on a circle gives a physical 
manifestation of this phenomenon and, when coupled with 
holomorphy, may be used to compute fractional instanton 
effects in four-dimensional gauge theories in the 
weakly coupled regime \cite{timmy,nicky}. Fractional 
instantons also appear in certain two-dimensional 
gauge theories and sigma-models, the most familiar 
example being the ${\cal N}=(2,2)$, ${\bf CP}^{N-1}$ 
sigma-model. As for the four-dimensional case, compactification 
on a circle gives a new method to compute these effects 
using controlled, semi-classical techniques.

\newpage

\subsection*{Acknowledgements}

My thanks to John Brodie, 
Jerome Gauntlett, Neil Lambert, Adam Ritz and Paul 
Townsend for discussions on 
various issues, and to Luis Bettencourt for comments on the draft. 
I am supported by a Pappalardo fellowship, and 
thank Neil Pappalardo for his generosity. This work is supported 
in part by funds provided by the U.S. Department of Energy 
(D.O.E.) under cooperative research agreement 
\#DF-FC02-94ER40818.

\end{document}